\documentclass[final,3p,times]{elsarticle}
\usepackage{xspace}
\usepackage{amssymb}
\usepackage{amsmath}
\usepackage[hyphens]{url}
\usepackage{hyperref}
\newcommand{\PbPb}         {\mbox{Pb--Pb}\xspace}

\newcommand{\sNN}          {\ensuremath{\sqrt{s_{\mathrm{NN}}}}\xspace}
\newcommand{\pt}           {\ensuremath{p_{\rm T}}\xspace}

\newcommand{\yrange}[1]    {\mbox{$\left | y \right |~<~#1$}}

\newcommand{\vtwo}         {\ensuremath{v_{\rm 2}}\xspace}

\newcommand{\GeVc}         {\ensuremath{\mathrm{GeV}/c}\xspace}

\newcommand{\pipm}               {\ensuremath{\pi^{\pm}}\xspace}

\newcommand{\DzerotoKpi}        {\ensuremath{{\rm D}^0 \to {\rm K}^-\pi^+}\xspace}
\newcommand{\DplustoKpipi}      {\ensuremath{{\rm D}^+\to {\rm K}^-\pi^+\pi^+}\xspace}

\newcommand{\DstophipitoKKpi}   {\ensuremath{{\rm D_s^{+}\to \phi\pi^+\to K^-K^+\pi^+}}\xspace}

\newcommand{\Dzero}             {\ensuremath{\mathrm{D^0}}\xspace}

\newcommand{\Dplus}             {\ensuremath{\mathrm{D^+}}\xspace}

\newcommand{\Ds}                {\ensuremath{\mathrm{D_s^+}}\xspace}
\newcommand{\Lc}                {\ensuremath{\Lambda_\mathrm{c}^+}\xspace}

\newcommand{\LctopKpi}          {\ensuremath{\Lambda_\mathrm{c}^+\to\mathrm{pK^-\pi^+}}\xspace}

\journal{Journal of Subatomic Particles and Cosmology}

\begin{document}

\begin{frontmatter}

\title{Charm-quark collectivity from small to large systems with ALICE}

\author{Marcello Di Costanzo\textsuperscript{a,b} for the ALICE Collaboration}
\affiliation[uni]{organization={Polytechnic University of Turin},
             addressline={Corso Duca degli Abruzzi 24},
             city={Turin},
             postcode={10129},
             state={Piedmont},
             country={Italy}}

 \affiliation[research]{organization={INFN Turin},
             addressline={Via P. Giuria 1},
             city={Turin},
             postcode={10125},
             state={Piedmont},
             country={Italy}}

\begin{abstract}
  In these proceedings, the elliptic flow (\vtwo) measurement of prompt charm hadrons 
  in lead--lead (Pb--Pb) and oxygen--oxygen (OO) collisions at $\sqrt{s_{\mathrm{NN}}} = 5.36$ TeV, 
  using the data collected during LHC Run 3 by the ALICE detector, is presented. 
  The analysis is performed at midrapidity ($|y| < 0.8$) and hadronic decay channels are used to reconstruct the signal candidates.
  The \vtwo coefficient is measured via the Scalar Product (SP) technique. 
  The \vtwo of prompt \Dzero, \Dplus, \Ds mesons, and \Lc baryons in semicentral Pb--Pb collisions is reported. 
  These results achieve unprecedented precision and low-$\pt$ reach for the D mesons and highlight the first observation of the splitting 
  of baryon and meson elliptic flow at intermediate $\pt$ in the charm sector.
  The prompt D-meson \vtwo is also measured in peripheral Pb--Pb collisions and central OO 
  collisions, probing the degree of thermal equilibration reached by charm quarks propagating in progressively smaller quark--gluon plasma media.
\end{abstract}


\end{frontmatter}



\section{Introduction}
The quark--gluon plasma (QGP) is a color-deconfined state of strongly interacting matter, predicted by lattice 
Quantum Chromodynamics (QCD) to exist at low baryochemical potential and temperatures above 155 MeV~\cite{Ratti:2018ksb}. 
Its properties can be studied in ultrarelativistic ion collisions at the 
Large Hadron Collider (LHC), where a strongly interacting medium is formed and 
only persists for timescales up to the order of 10~fm/$c$.
Due to their large mass, heavy quarks are produced in hard scatterings occurring before the QGP formation 
and exchange energy and momentum with the medium constituents throughout the full evolution of the system. 
These interactions modify the production yields and angular 
distributions in the azimuthal plane of final-state hadrons, making them sensitive probes of QGP properties 
and of the mechanisms driving hadron formation. 
Given the non-perturbative nature of QCD in this regime, 
such phenomena are primarily described by phenomenological transport models. 
In this framework, experimental measurements aim at constraining the spatial diffusion coefficient $D_\text{s}$, 
which is connected to the fundamental properties of the QGP.\\

The interactions of quarks with the QGP medium can be studied through the elliptic flow observable, which 
develops as the generated medium featuring partonic degrees of freedom expands under non-isotropic pressure gradients. 
In non-central collisions, the almond-shaped overlap region of the colliding nuclei is understood 
to be the origin of elliptic flow. 
For increasingly central collisions and for small colliding nuclei, the driving mechanisms become the fluctuations 
in the positions of the nucleons inside the colliding nuclei.
A modulation of the particle yields with respect to the reaction plane (RP) containing the impact parameter 
of the collision and the beam axis is observed, which can be described via a Fourier series expansion
\begin{equation}
    \frac{\text{d}N}{\text{d}\varphi} = \frac{N_0}{2\pi} \bigg( 1 + 2 \sum_{n=1}^{\infty} v_{n} (\cos[n(\varphi - \Psi_{\text{RP}})]) 
    \bigg), \qquad v_n = \langle \cos[n(\varphi - \Psi_{\text{RP}})] \rangle,
\end{equation}
where $\varphi$ indicates the azimuthal angle of the emitted particle, $\Psi_{\text{RP}}$ the azimuthal angle 
of the reaction plane and \vtwo represents the \textit{elliptic flow coefficient}.  
The hadronization mechanisms of charm hadrons influence the measured \vtwo values. 
In case of production by fragmentation of a charm quark, the \vtwo reflects the interactions of the charm 
quark in the QGP medium. 
In case of production via coalescence, the \vtwo 
also includes the contribution from the interactions of the captured light quarks.
Lastly, when the hadron, referred to as \textit{non-prompt} in this case, is produced in the decay of a hadron with beauty-quark content, its 
\vtwo reflects the interactions of the beauty quark in the QGP medium, with a slight smearing due to the decay kinematics. 
The corresponding \vtwo is expected to be smaller in this case due to the large mass difference between charm and beauty quarks. 
Finally, the elastic and inelastic rescattering processes occurring in the hadron gas phase before the freeze-out can further contribute to the 
measured \vtwo.\\

The presented results use the data sample of Pb--Pb collisions recorded by ALICE in 2023 ($\mathcal{L}_{\text{int}} 
\approx 1.5$ nb$^{-1}$) and of OO collisions recorded in 2025 
($\mathcal{L}_{\text{int}} \approx 5$ nb$^{-1}$).
The ITS, TPC, and TOF detectors reconstruct charged particles at midrapidity ($|y|<0.8$) and provide particle identification (PID) 
information, with improved performance compared to LHC Run~2 owing to major upgrades implemented during LHC Long Shutdown~2~\cite{ALICE:2023udb}.
Charm hadrons are reconstructed in selected hadronic decay channels along with their charge 
conjugates: $\DzerotoKpi$, $\DplustoKpipi$, $\DstophipitoKKpi$, and $\LctopKpi$. 
The displaced decay-vertex topologies determined by the lifetime of charm and beauty hadrons, of the order of 100 $\mu$m and 400 $\mu$m 
respectively, along with the PID information of their decay products are used as input for multi-class Boosted Decision Tree (BDT)-based machine learning models  
to separate the background, prompt, and non-prompt contributions. 
The Scalar Product (SP) method~\cite{Poskanzer:1998yz} is employed to extract the \vtwo values. 
Since the prompt and non-prompt charm-hadron \vtwo coefficients differ and exclusive samples of each contribution cannot be obtained from 
BDT selections, a data-driven procedure is employed to correct for the non-prompt contamination and extract the prompt \vtwo values. 
This strategy is reported in detail in Ref.~\cite{ALICE:2026zcz}.

\begin{figure}[bt]
    \centering
    \includegraphics[width=0.41\textwidth]{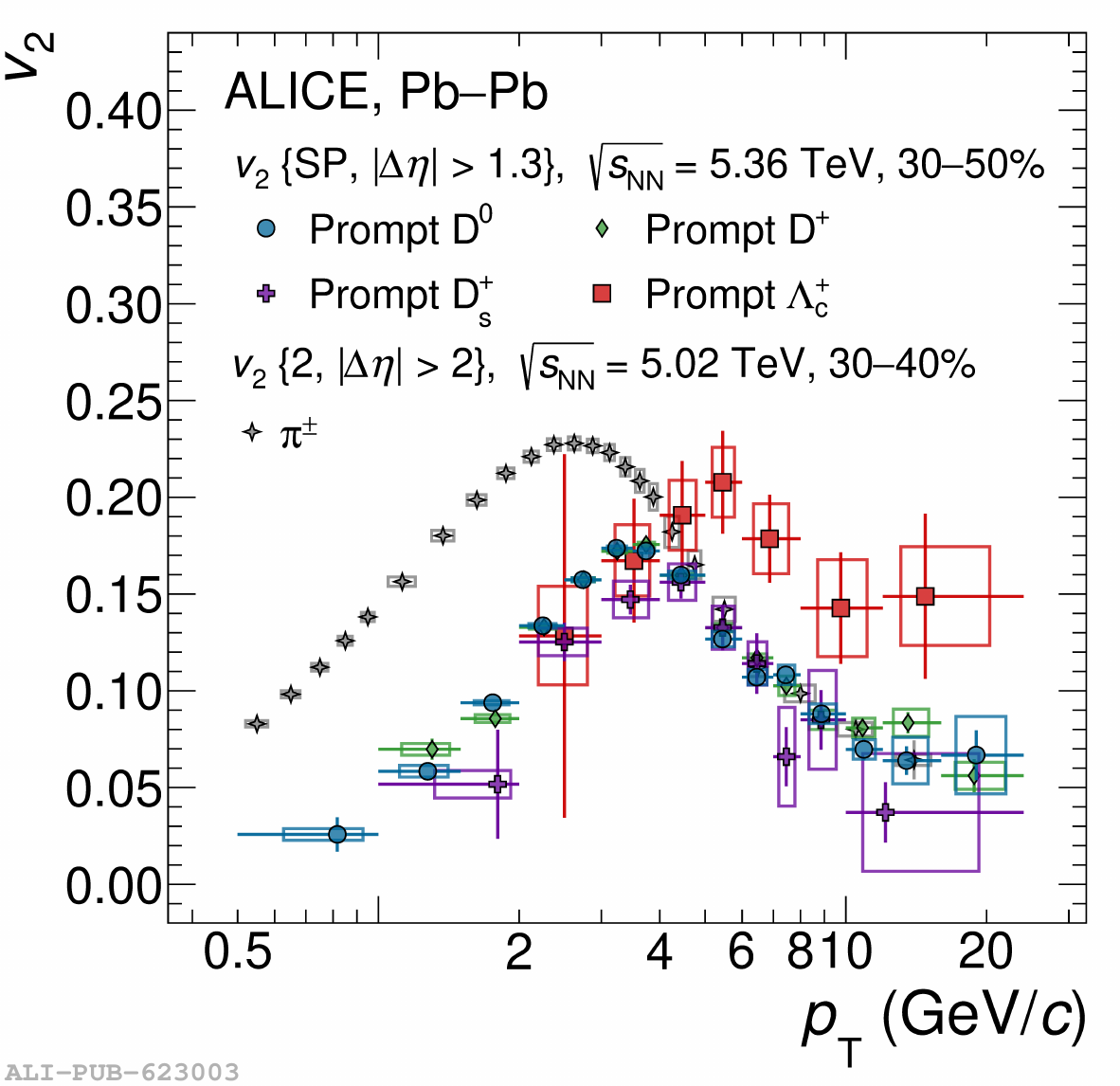}
    \includegraphics[width=0.41\textwidth]{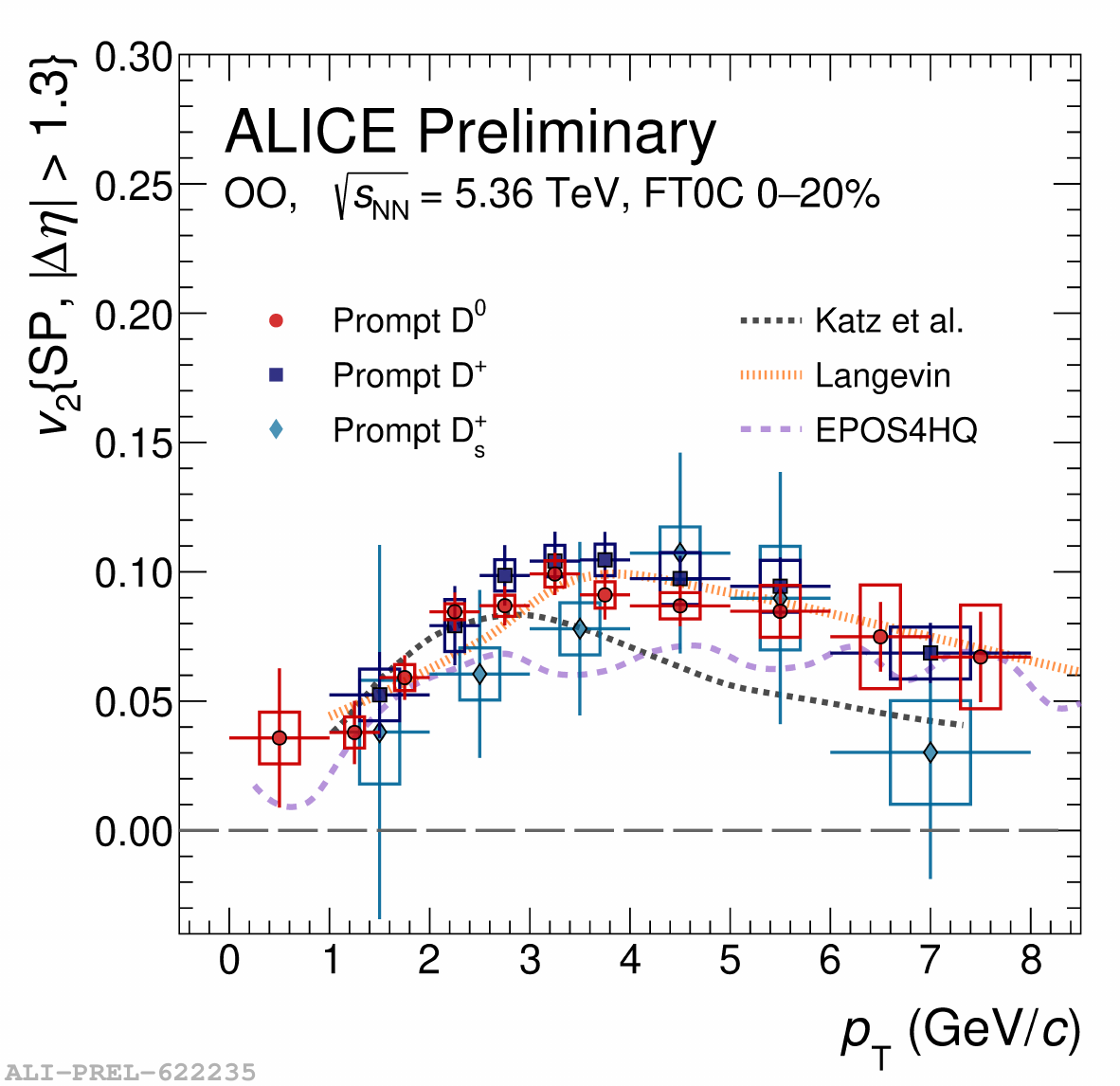}
    \caption{Left: measurement of prompt \Dzero-, \Dplus-, \Ds-meson and \Lc-baryon \vtwo in semicentral Pb--Pb collisions at $\sqrt{s_{\rm NN}} = 5.36$ TeV, 
    taken from Ref.~\cite{ALICE:2026zcz}, compared with pions and $\Lambda$ baryons from Ref.~\cite{ALICE:2018yph}. 
    Right: measurement of prompt \Dzero-, \Dplus-, and \Ds-meson \vtwo in central OO collisions at $\sqrt{s_{\rm NN}} = 5.36$ TeV,  
    compared with transport model predictions~\cite{Katz:2019qwv, Li:2021xbd, Zhao:2024oma}.}
    \label{fig:v2_dmesons_lc_pbpb}
\end{figure}

\section{Results}
The elliptic flow of prompt \Dzero, \Dplus, \Ds mesons and \Lc baryons at midrapidity (\yrange 0.8) in \PbPb collisions at $\sNN = 5.36$~TeV in 
the 30--50\% centrality class as a function of \pt is shown in Fig.~\ref{fig:v2_dmesons_lc_pbpb}. 
In the same figure, the \vtwo of charged pions and $\Lambda$ baryons (\yrange 0.5) 
measured at $\sNN = 5.02$~TeV in the 30--40\% centrality class~\cite{ALICE:2018yph} is shown. 

A positive \vtwo is measured for all charm-hadron species over the covered \pt interval, and the \vtwo for \Dzero mesons is measured for \pt 
lower than $1~\GeVc$ for the first time. The prompt \Dzero- and \Dplus-meson \vtwo are compatible within the experimental uncertainties. At low 
transverse momentum ($\pt <4~\GeVc$), the charm-hadron \vtwo increases with \pt and exhibits a clear mass ordering ($\vtwo(\pipm) > \vtwo(\Lambda) > \vtwo(\rm D)$). 
This behavior, already observed in heavy-ion collisions for light- and heavy-flavor hadrons~\cite{ALICE:2018yph}, highlights the interplay between radial and 
elliptic flow within a hydrodynamically expanding medium. In the presence of a common 
velocity field, heavier particles receive a larger momentum boost than lighter ones. This modifies the momentum spectra of the produced particles, 
leading to a shift of their \vtwo distribution toward higher \pt, and ultimately causing the observed mass ordering. 
The measured \vtwo values indicate a lower elliptic flow of \Ds mesons compared to \Dzero mesons with a 
$2.6\sigma$ significance. 
Ref.~\cite{He:2012df} suggests an early kinetic freeze-out of strange relative to non-strange hadrons during 
the hadronic phase of the collision, which would reduce the amount of flow transferred to the \Ds through multiple low-energy interactions with the 
hadronic gas. An alternative interpretation could be related to a sequential hadronization 
scenario~\cite{Xu:2025ivv}, in which strange D mesons are expected to coalesce on a shorter timescale than non-strange ones, 
resulting in a shorter propagation time in the QGP and a smaller \Ds \vtwo compared to \Dzero. The present uncertainties 
preclude a definitive conclusion, but the analysis of the 
full ALICE Run 3 Pb--Pb dataset is expected to clarify this point.\\

For $\pt > 4$~\GeVc, the \vtwo of D mesons smoothly decreases with increasing \pt, following the trend and magnitude of the \pipm \vtwo. In the 
$4 < \pt < 12~\GeVc$ interval, the \vtwo of \Lc baryons is larger than that of \Dzero mesons with a $3.7\sigma$ significance
This behavior represents the first evidence that the baryon--meson \vtwo splitting observed for light-flavor hadrons~\cite{ALICE:2018yph} 
extends to the heavy-flavor sector, suggesting a common underlying mechanism across different quark flavors. The observed \vtwo splitting 
between baryons and mesons is commonly interpreted as a consequence of hadron formation via recombination or coalescence in a medium 
featuring partonic degrees of freedom. 
For \pt higher than 10~\GeVc, the \vtwo values are compatible within 
uncertainties and remain positive due to a common energy loss experienced 
by high-momentum partons. \\

\begin{figure}[bt]
    \centering
    \includegraphics[width=0.41\textwidth]{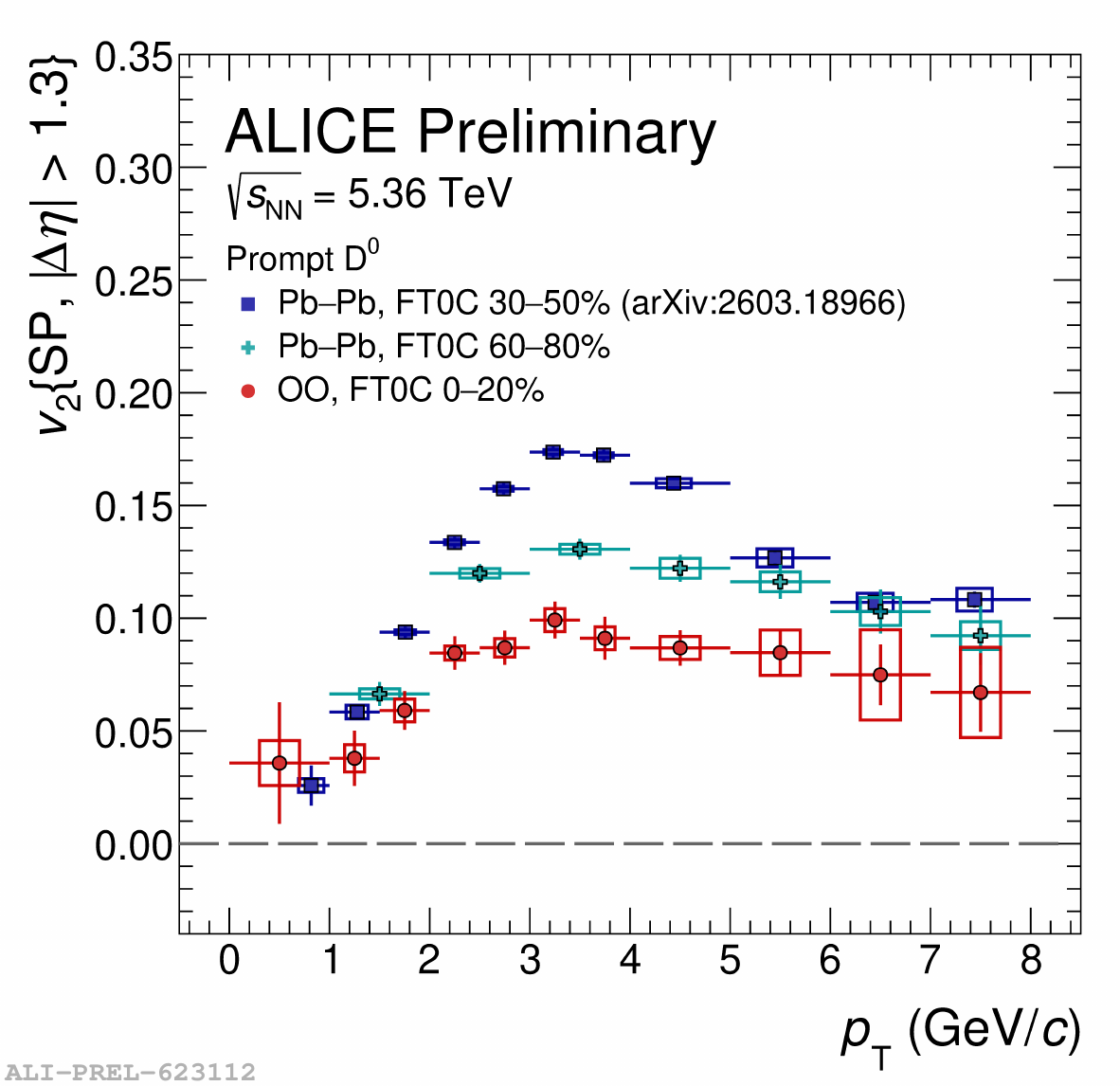}
    \includegraphics[width=0.41\textwidth]{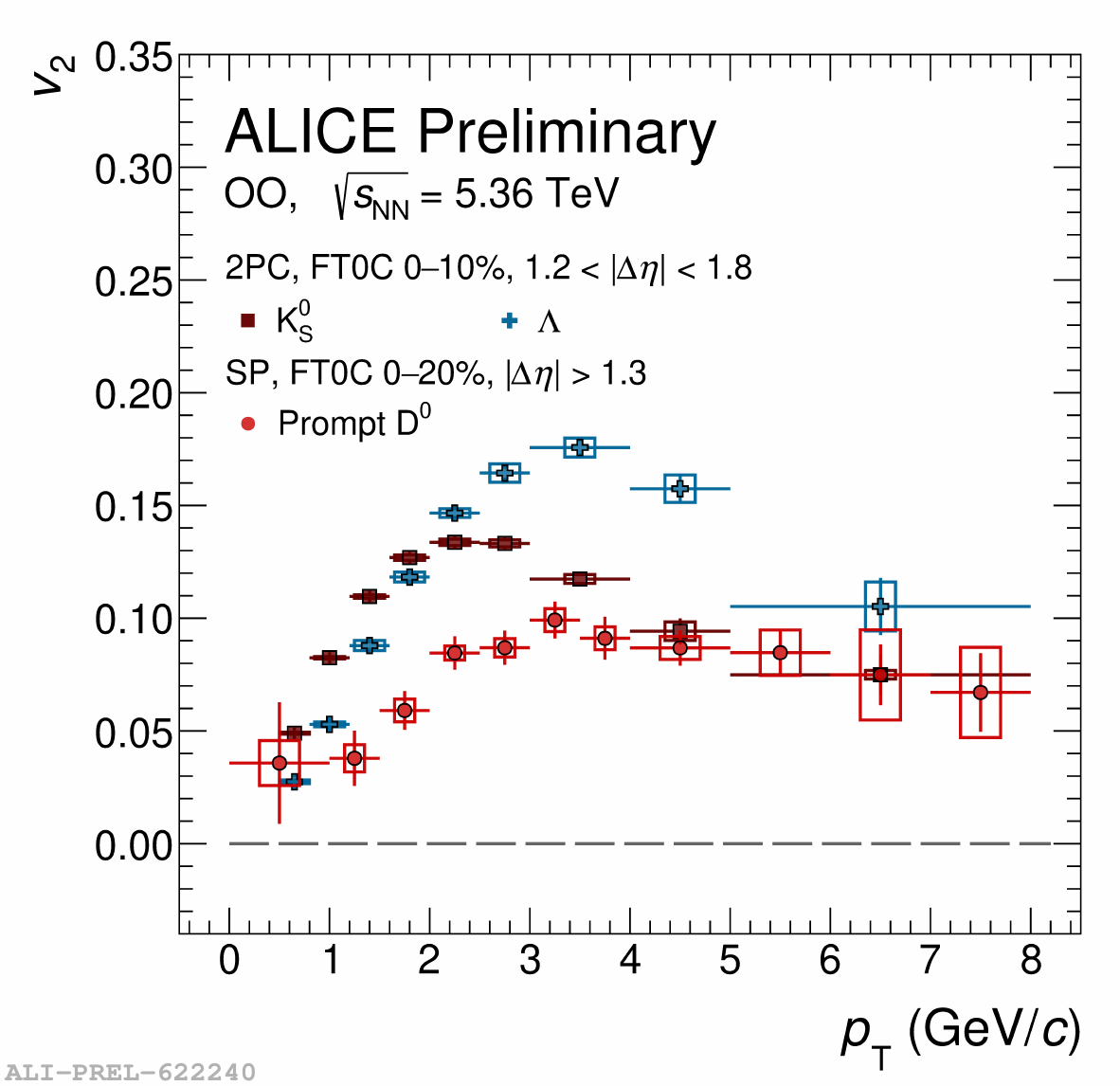}
    \caption{Left: comparison of prompt \Dzero-meson \vtwo in semicentral, peripheral Pb--Pb and central OO collisions. 
    Right: comparison of \vtwo of prompt \Dzero, K$_{\text{S}}^0$ mesons and $\Lambda$ baryons 
    in central OO collisions.}
    \label{fig:v2_dmesons_oo}
\end{figure}

The right panel of Fig.~\ref{fig:v2_dmesons_lc_pbpb} reports the first measurement of the prompt 
\Dzero-, \Dplus-, and \Ds-meson \vtwo at midrapidity (\yrange 0.8) in central OO collisions (0--20\% 
centrality class) at $\sNN = 5.36$~TeV. 
The \vtwo of \Dzero, \Dplus, and \Ds mesons are compatible within the uncertainties and exhibit 
positive values over the considered $\pt$ range. 
Transport models including QGP features~\cite{Katz:2019qwv, Li:2021xbd, Zhao:2024oma}, 
also shown in the same panel, are able to qualitatively reproduce the observed 
features. 
Since charm quarks are expected to develop elliptic flow through interactions with the expanding medium 
during the later stages of the QGP evolution, this measurement probes the charm-quark degree of thermalization 
achieved in a QGP medium that is smaller and shorter-lived than that produced in Pb--Pb collisions. 
The left panel of Fig.~\ref{fig:v2_dmesons_oo} compares the prompt \Dzero-meson \vtwo measured in central 
OO, semicentral Pb--Pb, and peripheral Pb--Pb collisions. While the \pt dependence is similar in the 
three collision systems, with \vtwo increasing up to \pt~$\approx$~3.5~\GeVc and decreasing at higher 
transverse momentum, the \vtwo magnitude progressively decreases from semicentral Pb--Pb to 
peripheral Pb--Pb and finally to central OO collisions. The smaller \vtwo observed in central 
OO and peripheral Pb--Pb collisions relative to semicentral Pb--Pb collisions reflects the reduced size and 
lifetime of the produced QGP medium. The remaining difference between central OO and 
peripheral Pb--Pb collisions, which have comparable particle multiplicities, is instead mainly sensitive 
to the initial geometry, since peripheral Pb--Pb collisions feature a larger eccentricity. Together, 
these complementary measurements help disentangle the effects of medium size and initial geometry on 
charm-hadron collectivity, providing more stringent constraints on charm-quark transport in the QGP.\\

The right panel of Fig.~\ref{fig:v2_dmesons_oo} compares the measurement of the prompt \Dzero-meson \vtwo in central OO collisions with the measurement of 
the $\Lambda$-baryon and K$^0_{\text{S}}$-meson \vtwo in the 0--10\% centrality class using the two-particle correlation analysis technique (2PC). 
In the low \pt region below 3~\GeVc, the \vtwo values of the three hadron species are ordered according 
to their mass. This is consistent with the mass ordering effect which is expected from a hydrodynamic picture of the evolution of the collision. 
In the intermediate \pt region (3--5~\GeVc), a deviation of the prompt \Dzero-meson \vtwo is observed with respect to the $\Lambda$-baryon \vtwo, 
favoring the scenario in which a meson-baryon splitting occurs. At high \pt (5--8~\GeVc), the \vtwo values of the three hadron 
species are compatible within the uncertainties, suggesting that in this kinematic range the mechanism driving the elliptic flow formation is 
the path-length dependence of the energy loss inside a QGP medium.

\section{Conclusion}
These proceedings report the most precise measurement of prompt charm-meson \vtwo down to low \pt and first 
measurement of the prompt $\Lc$-baryon \vtwo in semicentral Pb--Pb collisions. 
The results provide evidence for the meson–baryon splitting of the \vtwo observable in the charm sector, 
and hint at a lower \vtwo for $\Ds$ mesons than for non-strange D mesons 
for $\pt<4$~\GeVc, although the current uncertainties do not support a firm conclusion. 
The prompt \vtwo is also measured for the first time in peripheral Pb--Pb and central 
OO collisions, where a smaller and shorter-lived QGP phase is produced. 
These measurements probe QGP media of complementary geometrical properties, providing inputs to disentangle 
the size and eccentricity contributions to the charm-hadron \vtwo.
The D-meson \vtwo is positive in both cases, indicating that at least a partial thermal equilibration 
has occurred. For the OO results, the comparison with \vtwo measurements in the light-flavor sector 
reveals the characteristic mass ordering at low \pt and baryon--meson grouping at intermediate \pt, 
consistent with QGP formation.

\bibliographystyle{elsarticle-num}
\bibliography{sqm2026_proceedings}

\end{document}